\documentclass[11pt]{amsart}
\usepackage[foot]{amsaddr}
\usepackage{fullpage}
\usepackage{amsfonts,amsmath,amsthm,array,amssymb}
\usepackage{graphicx}
\usepackage[font=footnotesize]{caption}

\usepackage[version=2]{mhchem}
\newcommand{\n}{\textrm{n} }
\newcommand{\ie}{i.e.,\ }
\newcommand{\eg}{e.g.,\ }
\mathchardef\minus="002D
\newcommand{\conc}[1]{\left[ #1 \right]}

\newcommand{\Erk}{\textrm{Erk}}

\newcommand{\Erkp}{\textrm{Erkp}}
\newcommand{\Mek}{\textrm{Mek}}
\newcommand{\Merk}{\textrm{M-Erk}}
\newcommand{\field}[1]{\mathbb{#1}} 
\newcommand{\rmnum}[1]{\romannumeral #1}
\begin{document}
\title{Dependence of MAPK mediated signaling on Erk isoforms and differences in nuclear shuttling}

\author{Heather A. Harrington$^{*1}$, Micha{\l} Komorowski$^1$, Mariano Beguerisse D\'iaz $^{2}$, \\
Gian Michele Ratto $^{3}$, and Michael P.H. Stumpf$^1$}
\address{$^*$ Corresponding Author}
\address{$^1$ Division of Molecular Biosciences, Imperial College London}
\address{$^2$ Division of Biology and Department of Mathematics, Imperial College London}
\address{$^3$ Istituto Nanoscienze and Scuola Normale Superiore, Pisa, Italy}
\email{heather.harrington06@imperial.ac.uk}

\maketitle



\begin{abstract}The mitogen activated protein kinase (MAPK) family of proteins is involved in 
regulating cellular fate activities such as proliferation, differentiation and 
apoptosis.  Their fundamental importance has attracted considerable attention on different aspects of the MAPK signaling dynamics; this is particularly true for the Erk/Mek system, which has become the canonical example for MAPK signaling systems. Erk exists in many different isoforms, of which the most widely studied are Erk1 and Erk2. Until recently, these two kinases were considered equivalent as they differ only subtly at the sequence level; however, these isoforms exhibit radically different trafficking between cytoplasm and nucleus. Here we use spatially resolved data on Erk1/2 to develop and analyze spatio-temporal models of these cascades; and we discuss how sensitivity analysis can be used to discriminate between mechanisms. We are especially interested in understanding why two such similar proteins should co-exist in the same organism, as their functional roles appear to be different. Our models elucidate some of the factors governing the interplay between processes and the Erk1/2 localization in different cellular compartments, including competition between isoforms. This methodology is applicable to a wide range of systems, such as activation cascades, where translocation of species occurs via signal pathways. Furthermore, our work may motivate additional emphasis for considering potentially different roles for isoforms that differ subtly at the sequence level. \\ \emph{Key words:} cell signaling, MAPK, Erk isoforms, sensitivity analysis
\end{abstract}


\section*{Introduction}

Cellular decision making processes require balanced and nuanced responses to environmental, physiological and developmental signals. Especially in eukaryotes this involves the interplay and concerted action of a number of molecular players, that receive external signals, broadcast them further into the cytoplasm, and, if a transcriptional response is called for, shuttle into the nucleus and activate the transcriptional machinery. The mitogen activated protein kinases (MAPK) are important relays in many of these signal transduction processes. They are, for example, involved in regulating cellular fates such as proliferation, differentiation and apoptosis \cite{muller:springer:2010}. The most widely studied MAPK, Erk1/2, is activated through phosphorylation by Mek, its MAPK kinase (MAPKK), and Mek in turn is activated by its cognate kinase; ultimately, the activator of the cascade of MAPK kinases is an initiator signal, such as Ras (MAPKK kinase). Activated kinases, such Erk or Mek, are deactivated via dephosphorylation by their respective phosphatases.
\par
In the active state Erk can shuttle into the nucleus where it can induce the relevant cellular responses. But continuous nucleo-cytoplasmic shuttling of Mek/Erk has been demonstrated even in the absence of activation \cite{volmat:jcs:2001,pouyssegur:biochempharm:2002,pouyssegur:ejb:2003,marchi:po:2008}.  Mek is believed to play a central role in the shuttling mechanism of Erk using its nuclear export sequence; however, due to its nuclear exclusion sequence, Mek is found mostly in the cytoplasm whereas Erk can accumulate in the nucleus \cite{marchi:po:2008,pouyssegur:ejb:2003}. 
\par
The two distinct isoforms of Erk, Erk1 and Erk2, appear to have distinctly different biological characteristics. Transgenic gene knockout mice lacking Erk2 (Erk2$^{-/-}$) result in embryonic lethality, whereas Erk1$^{-/-}$ are viable, fecund and suffer from only subtle deficiencies, \eg reduced T cell development \cite{mazzucchelli:neuron:2002,pages:science:1999}.  Erk2 can compensate in Erk1 deficient mice through as yet unclear mechanisms. Although Erk1/2 share approximately 85\% of the same amino acid sequence, both activated by Mek, and considered to behave similarly, shuttling of Erk1 from cytoplasm to nucleus is approximately three times slower than that of Erk2, regardless of whether the cell is starved or stimulated \cite{costa:jcs:2006,marchi:po:2008}. The rate of shuttling results from differences in the N-terminal domain of Erk1 and Erk2 \cite{marchi:bbrc:2010}. In addition to different trafficking speeds, Erk1 is approximately four times less abundant than Erk2 in NIH 3T3 cells \cite{lefloch:mcb:2008}. It is also believed that competition exists between Erk isoforms in binding to Mek \cite{vantaggiato:job:2006}.
\par
Activation cascades have been studied extensively, but given that experimental results often rely on incomplete or indirect observations of the underlying dynamics, and in light of sometimes partially contradictory experimental results, mathematical models have become increasingly important for understanding these cascades. Published models of activation cascades, such as MAPK, typically use a three tier structure, where the top tier is activated by a single phosphorylation, and the remaining two tiers are activated via double phosphorylation cycles. This structure can incorporate  positive feedback loops and the most widely studied model by Huang and Ferrell predicted ultrasensitivity in the MAPK signal \cite{ferrell:tbs:1996,ferrell:jbc:1997,huang:pnasu:1996}.  Their work initiated the development of several alternative MAPK cascade models, which exhibit cell switching \cite{xiong:n:2003,ferrell:tbs:1997,angeli:pnasusa:2004} and provide conditions for bistability (\eg via double phosphorylation through a distributive mechanism \cite{markevich:jcb:2004} or MAPKK sequestration \cite{legewie:bj:2007}).  Subsequent models have also included inhibition by the doubly phosphorylated MAPK onto the initial signal Ras, either directly or implicitly between cycles, via a negative-feedback loop. These models predicted sustained oscillations \cite{kholodenko:ejb:2000,ventura:pcb:2008}, which were then verified by further modeling and experimental work \cite{shankaran:msb:2009}. Additionally, other studies have found that parameter regions affect the dynamical behavior of the system which may yield multi-stability or oscillations \cite{qiao:pcb:2007,radhakrishnan:systbiol:2009,zumsande:jtb:2010}. Although many models resemble the form of the Huang--Ferrell model, many authors have explored the cascade in a linear manner \cite{heinrich:mc:2002,behar:bj:2007,qu:pb:2009} in order to study the effects of cascade length, signal transduction and adaptation, in the presence of MAPK specific feedback structures. 
\par
Shuttling may play a role in the activation/deactivation activity of Mek/Erk in both the nucleus and cytoplasm, which has motivated the recent development of spatio-temporal models \cite{fujioka:jbc:2006,radhakrishnan:systbiol:2009,shankaran:msb:2009,kholodenko:nrm:2010,kholodenko:wiley:2009}. A compartment model of MAPK using single phosphorylation and dephosphorylation loops \cite{fujioka:jbc:2006} was extended to include double phosphorylation of kinases and phosphatases which dephosphorylate the kinases \cite{shankaran:msb:2009}. Models have also been developed to study the differences between Erk isoforms \cite{schilling:msb:2009,marchi:po:2008,radhakrishnan:systbiol:2009}. A tiered MAPK model with Erk1, Erk2, Mek and upstream species of the MAPK cascade was developed identifying that Erk isoforms cross-regulate each other and may be linked to cell fate decisions \cite{schilling:msb:2009}. Another model focused on Erk1 dimerization using data from Erk1 wild type and the Erk1-$\Delta$4 mutant (unable to dimerize) and improved the nuclear transport model of \cite{fujioka:jbc:2006} by exploring different methods of phosphorylation \cite{radhakrishnan:systbiol:2009}. Finally, a simple model including only Erk1 and Erk2 species was constructed to study the nucleo-cytoplasmic shuttling speeds \cite{marchi:po:2008}. 
\par
The existence of two genetically nearly identical isoforms of Erk is hard to explain, especially in light of the above results. Here we seek to understand potential dynamical reasons for the existence of the Erk1 and Erk2 isoforms. To do so we have to consider their interactions with Mek and nuclear/cytoplasmic shuttling, which appear to be important in Mek/Erk mediated cellular decision making processes. We integrate these different processes into a single model and compare it to recent spatio-temporally resolved data and consider and contrast the behavior in two cell types, HeLa and NIH 3T3, and under different conditions.  We find that shuttling, \ie hitherto often ignored spatial aspects, exert great influence over the cellular response phenotypes and are able to provide a rationale for the existence of different Erk isoforms. Furthermore, our model supports the recently posed hypothesis  \cite{vantaggiato:job:2006} of the importance of competition between the Erk isoforms.

This study also lays the ground work for understanding a complex system in the biological sciences by developing an appropriate abstractions; in this case, through the development and analysis of a minimal model. In addition, these approaches are suitable in contexts where transport and/or different roles of molecules are unclear.

\section*{Materials and Methods}
\subsection*{A spatio-temporal model of Erk signalling} 
A reaction network of the spatio-temporal model is built around three species: $\Erk1$, $\Erk2$, and $\Mek$ are allowed to exist in different activation states, as complexes and in different cellular compartments (see Fig.~1A and find equations in Supplementary Material).  We assume Erk isoforms (Erk1/Erk2) exist in one of three states: inactive ($\Erk$), bound to Mek ($\Merk$), or active ($\Erkp$). Erk transitions from inactive to active through Mek whereby free phosphorylated Mek ($\Mek$) reversibly interacts with inactive $\Erk$. Upon Mek-Erk binding and formation of an intermediary complex ($\Merk$), Mek phosphorylates Erk resulting in activated $\Erk$ ($\Erkp$) and then dissociates from $\Erkp$. In order for $\Erkp$ to revert back to its inactive form, it must undergo dephosphorylation by a phosphatase. 
To denote the subcellular compartment of Erk, species are denoted by a subscript $n$ to indicate nuclear localization, \ie $\Erk1_{\n}$ is $\Erk1$ in the nuclear compartment. The biochemical reactions of the full system are given in Table~1 with a description in Table~3. Shuttling rates of Erk in HeLa cells have previously been established; however, to determine the shuttling of Erk species in NIH 3T3 cells, we performed FRAP experiments, as shown in Fig.~2, and these data enabled us to estimate transport rates. 

\begin{figure}
\begin{center}
\includegraphics[width=5in]{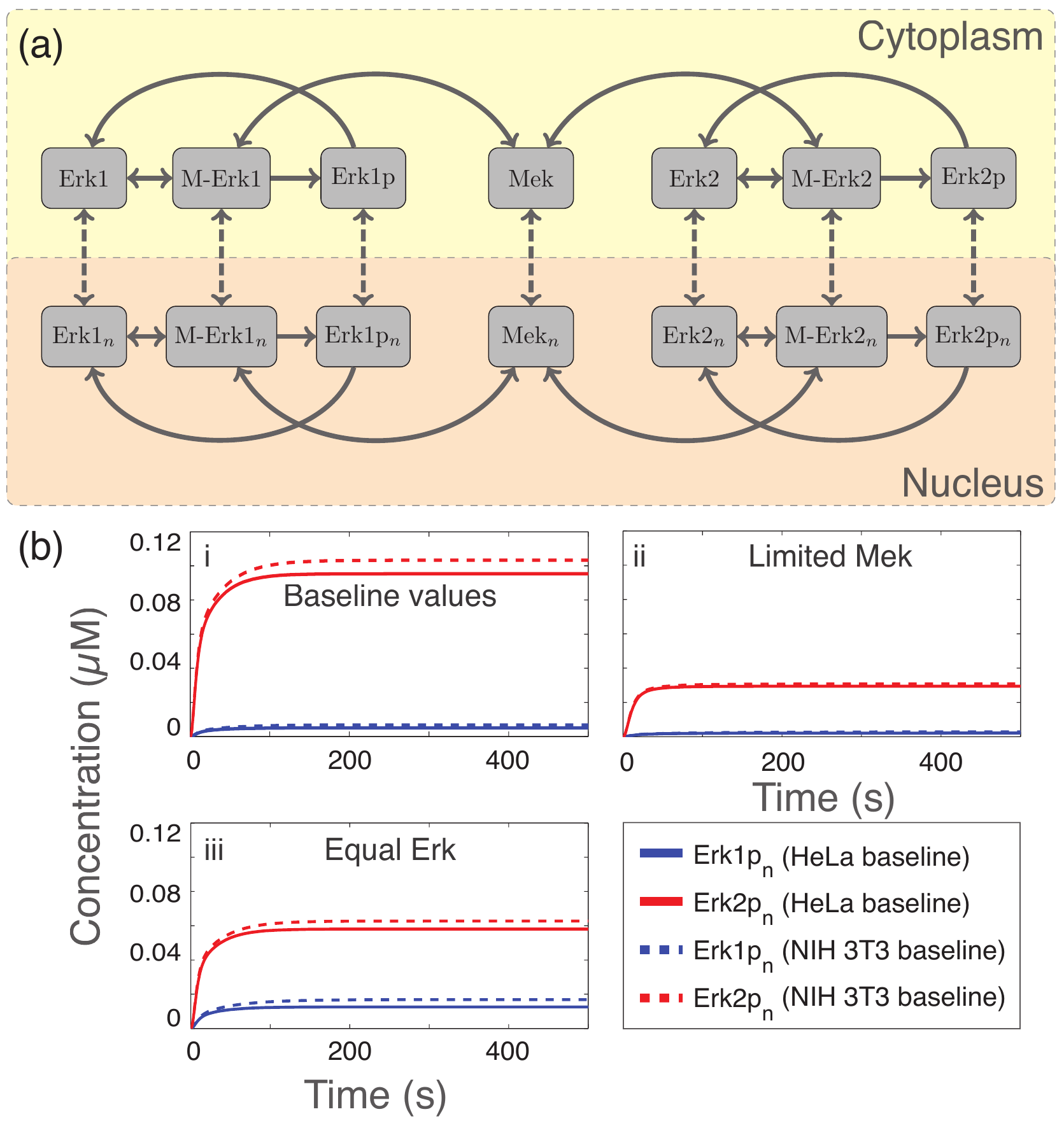}
\caption{Model construction. (a) Schematic of full Erk/Mek shuttling model. Inactive Erk can bind to Mek to form the M-Erk complex. Mek phosphorylates Erk into Erkp, then Mek dissociates and Erkp can be dephosphorylated to Erk. This behavior is symmetric for both Erk1 and Erk2 isoforms. These reactions (solid arrows) can occur in either the cytoplasm (yellow region) or nucleus (orange region) and all model species (gray boxes) can shuttle between compartments (dashed arrows). (b) Time course of nuclear $\Erk1p_{\n}$ and $\Erk2p_{\n}$ that reaches steady-state for HeLa and NIH 3T3 cells. (\rmnum{1}) Time course at baseline conditions. (\rmnum{2}) Time course at Limited Mek ($\Mek=0.2$). (\rmnum{3}) Time course at equal Erk1/2 ($\Erk1=\Erk2=0.5$).}
\end{center}
\end{figure}

\begin{figure}
\begin{center}
\includegraphics[width=4in]{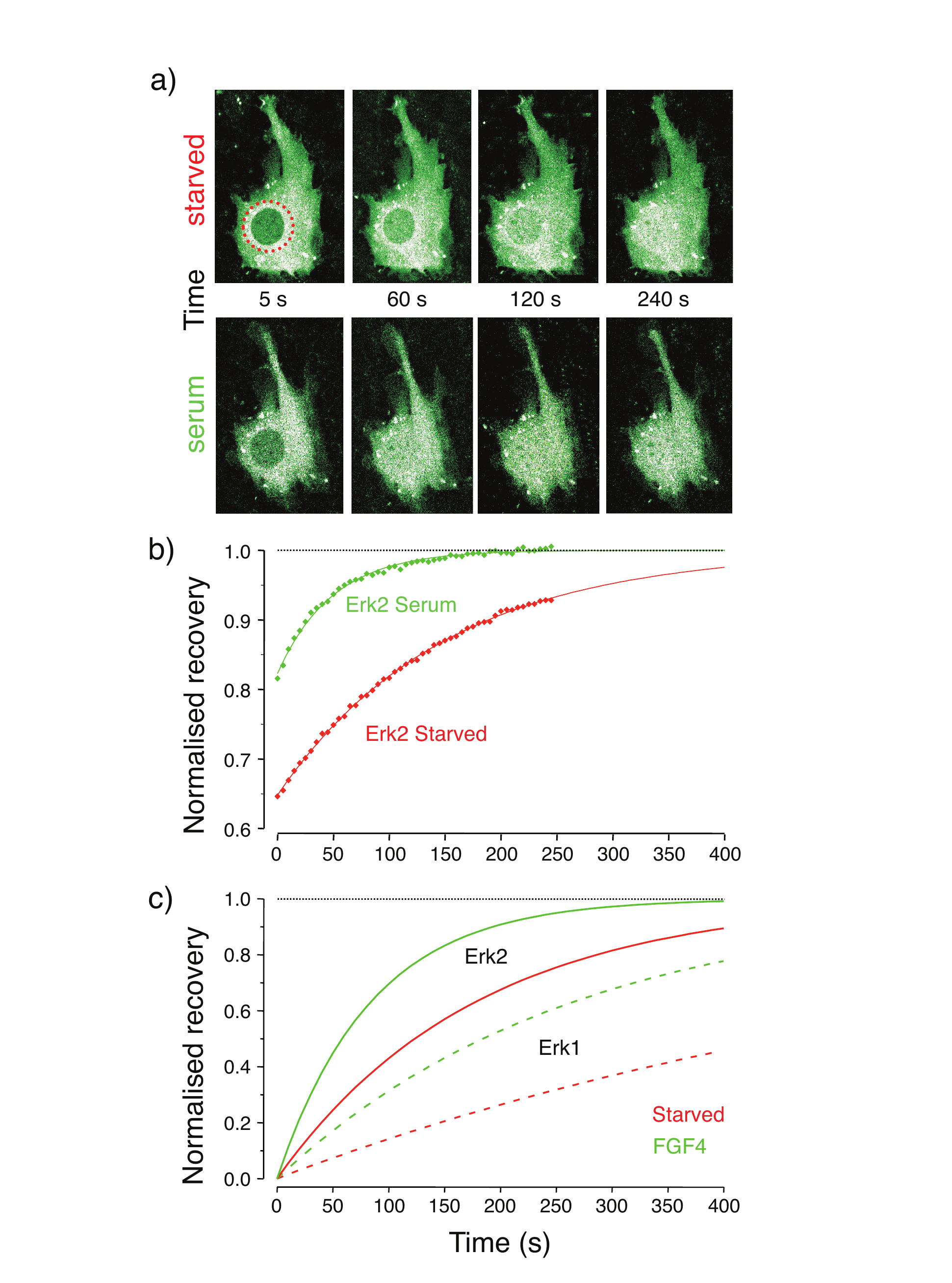}
\caption{Measure of the nucleus-cytoplasm trafficking of Erk-GFP  in living cells. a) Imaging of NIH 3T3 fibroblasts transfected with Erk2-GFP. Trafficking rates have been measured by analyzing the speed of recovery of fluorescence after bleaching of the nucleus (red dots in the first image): the faster is the recovery the higher is the rate of trafficking. Nuclear fluorescence was bleached at time 0 and the time lapse sequence shows its recovery. Cells were starved in 1\% fetal bovine serum for 24 hrs before performing the experiment. After compete recovery the cell were stimulated with 10\% serum and, after allowing 15 min for maximal Erk activation, a second bleach-recovery sequence was acquired. The calibration bar is 10 microns. b) Time course of the fluorescence recovery acquired for the illustrated cell before (red) and after activation of the Erk pathway (green). The continuous lines are fitting exponentials of time constants 152 and 49 s respectively. c) Erk1 and Erk2 have different trafficking rates: The four exponentials indicate the normalized time course of the fluorescence recovery for Erk1 (dotted lines) and Erk2 (continuous lines) in the starved condition (red) and after stimulation (green). Rates have been obtained from averaging rates measured in 30-60 cells in each condition (5).}
\end{center}
\end{figure}

\subsection*{Cell culture and transfection}
NIH 3T3 cells were plated on glass disks cultured in modified Dulbecco's medium supplemented with 10\% fetal bovine serum, and were transfected using Lipofectamine (GenePORTER 2, Genlantis, San Diego, CA).  To inactivate the ERK pathway, cells were starved by keeping them in 1\% FBS for 24 h before imaging. Further details and the design of the ERK-GFP fusion proteins are found in \cite{marchi:po:2008,costa:jcs:2006}.

\subsection*{FRAP measure of nucleus-cytoplasm shuttling}
Photo bleaching was preceded by the acquisition of a pre-bleach image that was used to estimate the loss of fluorescence due to bleaching and for data normalization. The nucleus of the cell was photobleached by repeated scans of the nucleus at high laser power. Bleaching was applied for approximately 8 s, which was sufficient to quench most of the nuclear fluorescence. Bleaching was followed by time-lapse acquisition to measure the recovery. Fluorescence was normalized by
$$F(t) = \frac{F_{\textrm{Tot}}}{F_{\textrm{Nuc}}} \cdot \frac{F_{\textrm{Nuc}} (f)}{F_{\textrm{Tot}} (t)}$$
where $F_{\textrm{Tot}}$ indicates the fluorescence (corrected for background) measured before bleaching in the entire cell and $F_{\textrm{Nuc}}$ indicates the same measure in the nucleus. This normalization corrects for bleaching caused by imaging. All imaging experiments were performed on a Olympus Fluoview 300 and Leica SL confocal microscopes.

\subsection*{Model parameterization}
We focus our analysis on  HeLA cells and mouse NIH 3T3 cells. Phosphorylation, shuttling rates and initial conditions have previously been specified for HeLa cells \cite{fujioka:jbc:2006,schoeberl:nb:2002} whereas shuttling rates here have been measured in mouse NIH 3T3 cells \cite{marchi:po:2008,costa:jcs:2006}. We use the kinetic phosphorylation and dephosphorylation parameters for an initial estimate for $\Erk2$ from the compartment model of $\Erk2$ in HeLa cells using a single phosphorylation loop \cite{fujioka:jbc:2006}.  $\Erk1$ phosphorylation was assumed to occur slightly faster than $\Erk2$ in primary erytheroid progenitor (CFU-E) cells \cite{schilling:msb:2009} and we adjusted the baseline parameter values for the complete model accordingly (Table~1).

\begin{table}[h]
        \begin{center} \caption{Reactions of Erk1/2 and Mek in full model. Each reaction highlights whether it is a forward or reversible reaction by the arrows as illustrated by solid arrows in Fig.~1A. All reaction rates are from \cite{fujioka:jbc:2006}. }
        {\small
        \begin{tabular}{|clcc|}
            \hline
            Number & Reaction & Forward rate ($\mu$M$^{-1}$ s$^{-1}$) & Reverse rate (s$^{-1}$)\\
            \hline\hline
            $1$ & $\Erk1 + \Mek \rightleftharpoons \Merk1$ & $8.8 \times 10^{-1}$ & $8.8 \times 10^{-2}$ \\
            $2$ & $\Merk1 \rightarrow \Mek + \Erk1p $ & $ 3 \times 10^{-1}$  &\\
            $3$ & $\Erk1_{\n} + \Mek_{\n} \rightleftharpoons \Merk1_{\n}$ & $8.8 \times 10^{-1}$  & $8.8 \times 10^{-2}$ \\
            $4$ & $\Merk1_{\n} \rightarrow \Mek_{\n} + \Erk1p_{\n}$ & $3 \times 10^{-1}$ & \\
            $5$ & $\Erk2 + \Mek \rightleftharpoons \Merk2$ & $8.8 \times 10^{-1}$  & $8.8 \times 10^{-2}$ \\
            $6$ & $\Merk2 \rightarrow \Mek + \Erk2p $ & $2 \times 10^{-1}$  &\\            
            $7$ & $\Erk2_{\n} + \Mek_{\n} \rightleftharpoons \Merk2_{\n}$ & $8.8 \times 10^{-1}$  & $8.8 \times 10^{-2}$ \\            
            $8$ & $\Merk2_{\n} \rightarrow \Mek_{\n} + \Erk2p_{\n}$ & $2 \times 10^{-1}$ & \\         
            $9$ & $\Erk1p  \rightarrow \Erk1$ & $1.4 \times 10^{-2} $  &\\            
            $10$ & $\Erk1p_{\n} \rightarrow \Erk1_{\n}$ & $1.4 \times 10^{-2}$ & \\  
            $11$ & $\Erk2p  \rightarrow \Erk2$ & $1.4 \times 10^{-2} $  &\\            
            $12$ & $\Erk2p_{\n} \rightarrow \Erk2_{\n}$ & $1.4 \times 10^{-2}$ & \\     
                         \hline
        \end{tabular}}       \end{center}
    \end{table}

Both HeLa and NIH 3T3 cells are 15-20 microns in diameter and have approximately the same volume. The import and export shuttling rates of species between the cytoplasm and nucleus include the ratio of cytoplasmic to nuclear volume as used in other compartment models \cite{fujioka:jbc:2006,shankaran:msb:2009,radhakrishnan:systbiol:2009}. We use the $\Erk2$ shuttling rates in HeLa cells from \cite{fujioka:jbc:2006} and then calibrated the shuttling rates to satisfy that $\Erk1$ shuttles three times slower than $\Erk2$ as seen in Fig.~2 and \cite{marchi:po:2008}. In starved cells there is no phosphorylation by Mek and primarily inactive $\Erk$ is shuttled; stimulated cells, on the other hand, (which have Mek/Erk activation) shuttle predominantly phosphorylated ($\Erkp$) species. The shuttling time constants ($\tau$) were calculated by first photobleaching the nucleus and then measuring the recovery of fluorescence of Erk in NIH 3T3 cells (see \cite{marchi:po:2008,costa:jcs:2006} for more details and Supplementary Material for transport estimation). Transport reactions and shuttling rates can be found in Table~2 and a description in Table~3.

 \begin{table}[h]
      \begin{center} \caption{Shuttling rates for HeLa and NIH 3T3 cells.Shuttling rates for HeLa and NIH 3T3 cells as shown dotted arrows in Fig.~1A. Parameters are estimated from \cite{fujioka:jbc:2006} and $^*$ denotes rate from \cite{marchi:po:2008}. } 
      {\small
        \begin{tabular}{@{}|cccccc|@{}}
            \hline
            & &  \multicolumn{2}{c}{Baseline HeLa cells} &  \multicolumn{2}{c|}{Baseline NIH 3T3 cells} \\ 
            Number & Transport & Export Rate (s$^{-1}$) & Import Rate (s$^{-1}$) & Export Rate (s$^{-1}$) & Import Rate (s$^{-1}$)\\
            \hline\hline
            $1$ & $\Erk1 \rightleftharpoons \Erk1_n$ &  $1.3 \times 10^{-2}$ &$4 \times 10^{-3}$ $^*$ &  2.1 $\times$ 10$^{-3}$ $^*$& 1.4 $\times$ 10$^{-3}$  $^*$\\
            $2$ & $\Mek \rightleftharpoons \Mek_n$ &  5.4 $\times$ 10$^{-1}$ &4 $\times$ 10$^{-2}$ &  5.4 $\times$ 10$^{-1}$ &4 $\times$ 10$^{-2}$   \\
            $3$ & $\Merk1 \rightleftharpoons \Merk1_n$ & 2.6 $\times$ 10$^{-1}$  & $1.17 \times 10^{-2}$ $^*$&  2.6 $\times$ 10$^{-1}$  & $1.17 \times 10^{-2}$ $^*$     \\
            $4$ & $\Erk1p \rightleftharpoons \Erk1p_n$ & $1.3 \times 10^{-2}$  &$4 \times 10^{-3}$ $^*$&    5.2 $\times$ 10$^{-3}$ $^*$& 4.7 $\times$ 10$^{-3}$  $^*$    \\
            $5$ & $\Erk2 \rightleftharpoons \Erk2_n$ & $1.8 \times 10^{-2}$  &$1.2 \times 10^{-2}$ &   7.8 $\times$ 10$^{-3}$ $^*$& 5.2 $\times$ 10$^{-3}$  $^*$     \\
           $6$ & $\Merk2 \rightleftharpoons \Merk2_n$ &  $2.6 \times 10^{-1}$ &$3.5 \times 10^{-2}$ &$2.6 \times 10^{-1}$ &$3.5 \times 10^{-2}$  \\
           $7$ & $\Erk2p \rightleftharpoons \Erk2p_n$ &  $1.3 \times 10^{-2}$  &$1.1 \times 10^{-2}$ & 1.65 $\times$ 10$^{-2}$ $^*$& 1.5 $\times$ 10$^{-2}$  $^*$        \\
             \hline
        \end{tabular}}
        \end{center}
    \end{table}

  \begin{table}[h]
        \begin{center} \caption{Description of model parameters.}
        {\small
        \begin{tabular}{|cl|}
            \hline
            Parameters & Description \\
            \hline\hline
            $k_1$ & forward interaction rate between cytoplasmic Mek and Erk1\\
            $k_{-1}$ & reverse interaction rate between cytoplasmic Mek and Erk1 \\
            $k_2$ &  phosphorylation rate of cytoplasmic Erk1  \\
            $k_3$ & forward interaction rate between nuclear Mek and Erk1  \\
            $k_{-3}$ & reverse interaction rate between nuclear Mek and Erk1 \\
            $k_4$ & phosphorylation rate of nuclear Erk1 \\
            $k_5$ & forward interaction rate between cytoplasmic Mek and Erk2\\
            $k_{-5}$ & reverse interaction rate between cytoplasmic Mek and Erk2  \\
            $k_6$ & phosphorylation rate of cytoplasmic Erk2 \\            
            $k_7$ & forward interaction rate between cytoplasmic Mek and Erk2 \\            
            $k_{-7}$ & reverse interaction rate between cytoplasmic Mek and Erk2  \\
            $k_8$ & phosphorylation rate of nuclear Erk2  \\         
            $k_9$ & dephosphorylation rate of cytoplamsic $\Erk1p$ \\            
            $k_{10}$ & dephosphorylation rate of nuclear $\Erk1p$  \\  
            $k_{11}$ & dephosphorylation rate of cytoplasmic $\Erk2p$ \\            
            $k_{12}$ & dephosphorylation rate of nuclear $\Erk2p$ \\     
            $s_1$ & import of Erk1 into the nucleus \\
            $s_{-1}$ & export of Erk1 out of the nucleus \\
            $s_2$ & import of Mek into the nucleus \\
            $s_2$ & export of Mek out the nucleus \\
            $s_3$ & import of M-Erk1 into the nucleus \\
            $s_{-3}$ & export of M-Erk1 into the nucleus \\
            $s_4$ & import of $\Erk1p$ into the nucleus \\
            $s_4$ & export of $\Erk1p$ out of the nucleus \\
            $s_5$ & import of Erk2 into the nucleus \\
            $s_{-5}$ & export of Erk2 out of the nucleus  \\
            $s_6$ & import of M-Erk2 into the nucleus \\            
            $s_{-6}$ & export of M-Erk2 out of the nucleus \\
            $s_7$ & import of $\Erk2p$ into the nucleus  \\            
            $s_{-7}$ & export of $\Erk2p$ out of the nucleus  \\
                         \hline
        \end{tabular}}        \end{center}
    \end{table}

 \begin{table}[h]
      \begin{center} \caption{Initial conditions of the model variables. Initial conditions of the model variables for different cell types.  Some species initial conditions may differ due to cell variability or cell type. Baseline values estimated from \cite{fujioka:jbc:2006,marchi:po:2008}.} 
      {\small
        \begin{tabular}{@{}|cccc|@{}}
            \hline
            & \multicolumn{3}{c|}{Initial concentration ($\mu$M)}  \\
            Species & Baseline & Equal Erk 1/2 & Limited Mek \\
            \hline\hline
            $\Erk1$ & $0.2$& $0.5$  & 0.2 \\
            $\Erk2$ & $0.8$  & $0.5$  & 0.8 \\
            $\Mek$ & $1$  & $1$  & 0.2 \\
             \hline
        \end{tabular}}
        \end{center}
    \end{table}

The non-zero baseline initial concentrations are the cytoplasmic species $\Erk1$, $\Erk2$, and $\Mek$. Estimates of initial levels of $\Erk$ in previous MAPK models of HeLa cells ranged between $0.96 - 2.1$ $\mu$M  \cite{schoeberl:nb:2002,fujioka:jbc:2006} and most previous models have initial $\Mek$ and $\Erk$ concentrations varying between $0.1 - 3$ $\mu$M \cite{fujioka:jbc:2006}. In HeLa cells, the ratio of Mek to Erk at the initial conditions is approximately 1.46 \cite{schoeberl:nb:2002,fujioka:jbc:2006}. We do not include inactive Mek in this model and we estimate that the ratio of active phosphorylated $\Mek$ to total Erk is one. Since Erk in previous models had been assumed to be approximately 1 $\mu$M \cite{schoeberl:nb:2002,fujioka:jbc:2006}  we assume the relation $\textrm{total~} (phosphorylated)~ \Mek \approx \textrm{total~} \Erk1 + \textrm{total~} \Erk2$, and the baseline initial concentrations of $\Erk$ also satisfy the condition that $\Erk1$ is four times less abundant than $\Erk2$ in NIH 3T3 cells \cite{lefloch:mcb:2008}. We also take into account that HeLa cells may have a different ratio of $\Erk1/\Erk2$ or different concentrations of active $\Mek$ either due to initial activation of the MAPK cascade or Mek sequestration effects, etc., than the baseline initial conditions as given in Table~4. 

\subsection*{Steady-state analysis}
We investigate the behavior of the system for the relevant baseline parameters and find that only one permissible and biologically relevant steady-state exists. Further exploration of the parameter space of the models using latin-hypercube sampling  \cite{mckay:tech:1979} confirms that there is only one biologically permissible steady-state across the biophysically plausible parameter space. This finding of the minimal model is in line with previous studies that looked into the classification of single and double phosphorylation cycle networks in terms of whether they yield multistability \cite{feliu:interface:2011}.

 \subsection*{Sensitivity analysis} To study the effects of the parameter values on the dynamics of  $\Erk1p_{\n}$ and $\Erk2p_{\n}$, we performed a detailed and comprehensive sensitivity analysis. We used a standard method based on the Fisher information \cite{rand2008mapping}, which assesses the change in system output as parameters are varied.  Suppose that we are interested in the values $\Erk1p_{\n}(t)$ and $\Erk2p_{\n}(t)$, denoting concentrations of nuclear $\Erk1p$ and $\Erk2p$ respectively, at times $t_{1}, . . ., t_{N}$. Then the sensitivity coefficient  for a parameter $k_{i}$ is 
\begin{equation}
S_{k_{j}}=\sum_{j=1}^{N} \left(\frac{\partial \Erk1p_{\n}(t_{j}) }{\partial k_{i} } \right)^{2}+ \left(\frac{\partial Erk2p_{\n} (t_{j})} {\partial t_{j} } \right)^{2}.
\end{equation}
As the trajectories $x_{\cdot}(t)$ are given by solutions of ordinary differential equations their derivatives $\frac{\partial x_{\cdot}(t) }{\partial k_{i} }$
can be easily calculated \cite{zwillinger:book:1998}. We have used this type of sensitivity coefficients as they have a clear information geometric interpretation as the infinitesimal distance in the state space, behave locally as the Kullback-Leibler divergence \cite{kass:stat-sci:1989}, and allow us to determine directions of trajectories change resulting from parameters changes.  Treating the initial conditions as parameters provides us with their sensitivities in an equally simple way \cite{rand2008mapping}.

\section*{Results}

Assuming a constant stimulus, as is the case for our experimental set-up, we simulate the model equations (see (Tables~1-4) for details) until steady-state. The time course of active nuclear $\Erk$ is given in Fig.~1a, and total nuclear $\Erk1$ and total nuclear $\Erk2$ abundances are shown in Fig.~S1a in the Supplementary Material. We take into account cell variability by studying the behavior at different initial concentrations of Erk and Mek within the known constraints; activation of the system is determined by the initial condition of active Mek.  Mek then activates $\Erk$, which can then traffic in and out of the nucleus; we are therefore interested in both the transient {\em activation time} (time until a species reaches steady-state) and the realised {\em steady-state} values of $\Erk1p_{\n}$ and $\Erk2p_{\n}$ (phopho-nuclear Erk isoforms).  We consider the realistic scenario of a smaller stimulus activating the system and take this into account by decreasing the initial level of activated Mek, which decreases the steady-state values of $\Erkp$ (Fig.~1b)). 

The overarching question is really why do nearly identical Erk 1/2 isoforms exist? In particular we seek to address this question by considering two related questions in more detail: (1) How does the behavior of these isoforms differ in different cell-types and under different physiological conditions? (2) How does nuclear shuttling differ between the two isoforms and how does this affect the overall nuclear activity levels of Erk? More generally, we develop a flexible mechanistic description of Erk mediated signaling that includes the nuclear translocation dynamics. This framework can be adapted to other molecules, \eg NFAT, Stat3/5 \cite{Timmer:2004ud} etc where we have distinct biochemical isoforms with different nuclear shuttling dynamics.

\subsection*{Minimal model}
To address the above questions, the behavior of Erk isoforms under varying physical conditions is more closely studied in the absence of transport (Fig.~3a). Sensitivity analyses identify $\Erk1p$ and $\Erk2p$ concentrations as most sensitive to the phosphorylation/dephosphorylation ($k_6/k_{11}$) rates of $\Erk1$ and $\Erk2$ (Fig.~3b). Analysis of the initial conditions also show that the available phosphorylated Mek increases in importance at lower signal strength (limited Mek) (Fig.~3b,c), which in an experimental context, is what we expect. The percentage of Mek that is usually phosphorylated depends on signal strength and this enzyme would greatly affect the steady state of Erkp. The balance between phosphorylation/dephosphorylation rates and initial conditions provides a setting for a competition scheme between Erk isoforms for its activator Mek; furthermore, available evidence shows that there is a close match between Mek and Erk1/2 concentrations. 
 
\begin{figure}
\begin{center}
\includegraphics[width=4in]{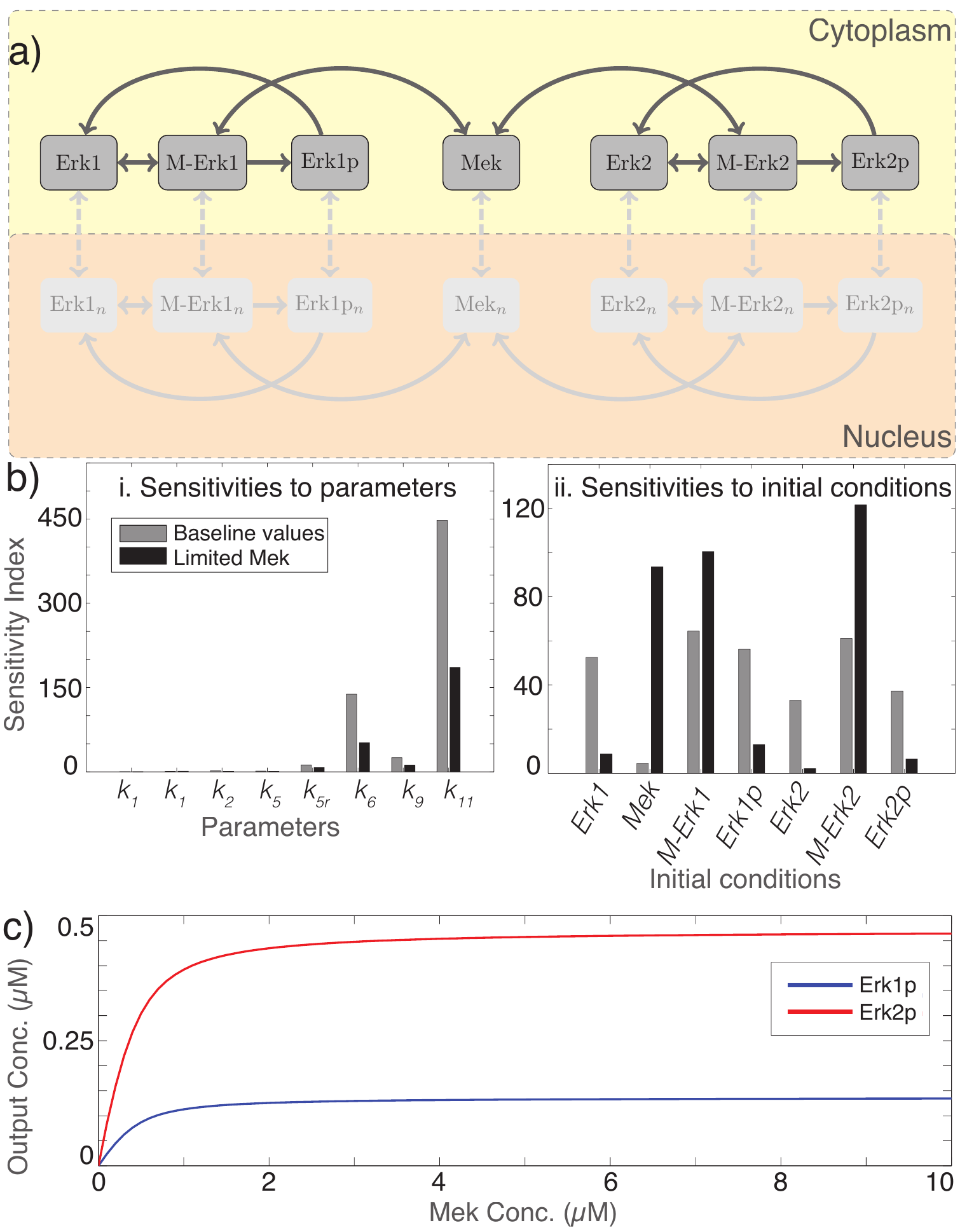}
\caption{Minimal Model. (a) Schematic of biochemical Erk/Mek model. (b) Sensitivity analysis for the biochemical model. Mean sensitivity index given by violet line (see Materials and methods). (\rmnum{1}) Sensitivity to parameters. (\rmnum{2}) Sensitivities to initial concentration.  }
\end{center}
\end{figure}

\begin{figure}
\begin{center}
\includegraphics[width=\textwidth]{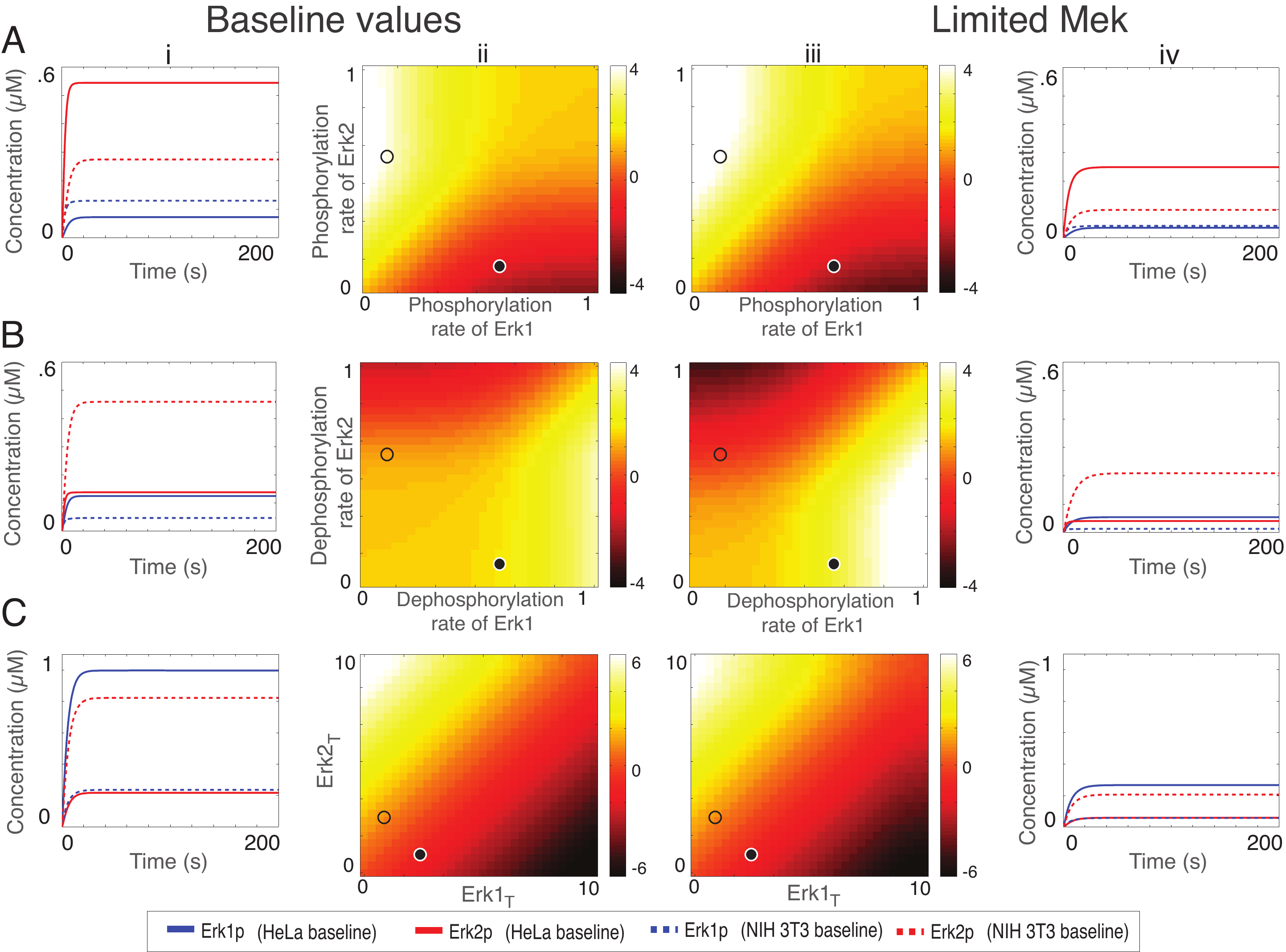}
\caption{Competition in minimal model at baseline and limited Mek (Mek=0.2) conditions. Competition scenarios are studied using heat maps (middle columns) and a time course (outside columns) of steady state. Heat maps (\rmnum{2}-\rmnum{3}) of parameter /initial condition values are given along the horizontal axis $(j)$ and vertical axis $(k)$ are varied and the associated color at each $(j,k)$-parameter/initial condition correspond to $\varphi=\log(\Erk2p/\Erk1p)$ where a positive value denotes more $\Erk2p$ than $\Erk1p$ and a negative value indicates more $\Erk1p$ than $\Erk2p$. Solid circle and open circle in heat map are shown in time course of $\Erk1p$ (blue) and $\Erk2p$ (red) in (\rmnum{1},~\rmnum{4}) where a solid circle corresponds to a solid line and an open circle corresponds to a dotted line. (a) Effects of phosphorylation rates on Erk1/2p. Values of ($k_2,k_6$) are varied, solid circle (solid line) is high $k_2=0.6,$ low $k_6 = 0.1$ and open circle (dotted line) is low $k_2=0.1, $ high $k_6 = 0.6$. (b) Effects of dephospho rylation rates on Erk1/2p. Values of ($k_9,k_{10}$) are varied, solid circle (solid line) is high $k_9=0.6,$ low $k_{10} = 0.1$ and open circle (dotted line) is low $k_9=0.1,$ high $k_{10} = 0.6$. (c) Effects of Erk1 and Erk2 initial conditions. Values of total Erk concentrations ($\Erk1_T$, $\Erk2_T$) are varied, solid circle (solid line) is high $\Erk1_T=2,$ low $\Erk2_T = 0.5$, and open circle (dashed line) is low $\Erk1_T=0.5,$ high $\Erk2_T = 2$. }
\end{center}
\end{figure}

\par
We investigate the effects of the phosphorylation/dephosphorylation rates on the relative abundances of $\Erk2p$ to $\Erk1p$ at baseline Mek and limited Mek conditions. By defining a log-scale indicator $$\varphi = \log \left( \frac{\Erk2p}{\Erk1p} \right),$$ the sign of $\varphi$ signifies the dominance of a specific phopho-Erk isoforms as physical conditions are changed. For example, given that the value of $\varphi$ identifies the change of phospho-Erk isoforms by orders of magnitude, a value of one indicates ten times more $\Erk2p$ than $\Erk1p$ whereas a negative value of $-1$ establishes there is ten times more $\Erk1p$ than $\Erk2p$ at steady state. The phosphorylation rates and dephosphorylation rates are each varied separately from 10$^{-3}$ to 10$^{1}$ in line with biophysical considerations and $\varphi$ is evaluated (Fig.~4a,b).  As the rate of phosphorylation of Erk1 increases, values of $\varphi$ become negative indicating dominance by $\Erk1p$ (see bottom right corner of Figs.~4a (\rmnum{2})), whereas an increase in the Erk2 phosphorylation rate demonstrates $\Erk2$ prevalence. The effect of the phosphorylation rate on $\varphi$ is asymmetric, meaning that there is a larger parameter space of values giving rise to Erk2p dominance (see non-overlapping curves in Fig.~4a (\rmnum{1})). Notably, the competition is exacerbated under limited Mek  conditions where the maximum/ minimum values of $\varphi$ are larger and smaller, respectively than the baseline conditions. Inspection of the dephosphorylation rates shows that an increase in dephosphorylation rate of Erk1 results in a larger positive $\varphi$, or an Erk2p steady-state bias, and vice versa for Erk2 dephosphorylation. This is also reflected in the heat map representation which reveals an asymmetry, with a bias towards $\Erk2p$, which is more   easily activated than $\Erk1p$. This is also apparent in the time course (Fig.~4b (\rmnum{1})) where equality between Erk1p and Erk2p steady state values exists for low Erk1 dephosphorlation rate, but not for low Erk2 dephosphorylation in Fig.~4b (\rmnum{1}). Under {\em limited Mek} conditions, both $\Erk1p$ and $\Erk2p$ steady-state values are much smaller and activation time increases  (Fig.~4a,b (\rmnum{4})). Moreover, for limited Mek conditions, the phosphorylation/dephosphorylation rate parameter space has a larger region for possible competition scenarios (see largest (white) and smallest (black) values of $\varphi$, corresponding to high Erk2p and Erk1p dominance in Fig.~4a,b (\rmnum{3})). This suggests that for certain phosphorylation/dephosphorylation rates, a limited stimulus would more strongly favor an $\Erk1p$ response than non-limited stimuli. We investigate how the total amount of $\Erk$ may vary across cells affects activation states (Fig.~4c) and observe that for small total $\Mek$, as well as baseline conditions, the initial amount greatly affects the steady-state value and it gives the expected result that as total Erk1 in the system increases, the total Erk1p also increases (Fig.~4c). Unlike the phosphorylation/dephosphorylation cases, however, the $\varphi$ indicator value at a given total (Erk1, Erk2) point does not change as Mek becomes limited (heat map indicator colors are nearly identical). 

The minimal model provides a number of insights, specifically that phosphorylation/dephosphorylation rates play an important role in the steady-state behavior of $\Erk1p$ and $\Erk2p$, whereas under limited Mek conditions, the parameter rate space suggests there is a stronger effect (larger $|\varphi |$) on the response. The value of the initial conditions can induce a $\Erk2p$ or $\Erk1p$ dominated response and limited Mek alters the steady-state value of this response. More generally, this model provides a simple framework for gaining insight into the components which control the competition between $\Erk1$ and $\Erk2$ for its kinase Mek, and we provide a indicator $\varphi$ for giving $\Erk1p$ or $\Erk2p$ cell response. 

\subsection*{Complete system}
The minimal model provides preliminary insights into the behavior of Mek and Erk isoforms. Even though the analysis of the minimal model is useful, further analysis of the complete system must be performed to acquire information about the role of translocation between the two cellular compartments (cytoplasm and nucleus). 

As shown in Fig.~5a, the equilibration dynamics and the steady state of $\Erk1p_{\n}$ and $\Erk2p_{\n}$ in HeLa and NIH 3T3 cells are most sensitive to the import shuttling rates of $\Erk1p$, $\Erk2p$, and $\Merk2$ ($s_4,s_7,s_3$); and the only sensitive biochemical reaction rate is the dephosphorylation rate of $\Erk2p_{\n}$ ($k_{12}$). The primary difference between the cell types is the sensitivity of Erk isoform shuttling speeds. Specifically, HeLa cells are more sensitive to Erk2p import shuttling whereas NIH 3T3 cells are more responsive to changes in Erk1p. Overall, both cell types are quantitatively most sensitive to the shuttling of Erk2. This analysis also highlights the similarities and differences in sensitivity between initial conditions of the minimal model and the complete system (Fig.~5b). The intermediary complex of Mek-Erk2 are sensitive in both models and the nuclear complexes are also sensitive. The drastic difference in steady state concentrations between the two cell conditions (limited Mek and baseline) that was observed in the minimal model does not transpire in the complete model. As shown in Fig.~5a,b, the sensitivity index of the most important parameters are nearly 100 times those of the initial conditions. Investigating the stimulus-response curve reveals that unlike the minimal model $\Mek(0)=1$ does not saturate $\Erk2p_{\n}$ and requires a stronger stimulus (compare Fig.~3c and Fig.~5c). The sensitivity analysis was repeated for limited Mek conditions, which did not present any difference in the importance of parameters or initial conditions.

Given the importance of the shuttling rates on $\Erk1p_{\n}$ and $\Erk2p_{\n}$ concentrations, we investigate the response of HeLa and NIH 3T3 cells to different conditions affecting shuttling.  For both cell types, we find that increasing the Erk1p shuttling rates (Fig.~6a), Erk2p shuttling rates (Fig.~6b), and Mek shuttling (Fig.~6c) by an order of magnitude increases the steady-state and the activation time. While the sensitivity analysis suggests that NIH 3T3 cells are more sensitive to Erk1p shuttling, the differences between Fig.~6a (\rmnum{1}-\rmnum{2}) are slight. In NIH 3T3 cells, the activation time of $\Erk1p_{\n}$ is shorter and the steady state is slightly larger. As shown in Fig.~6a,b, Erk1p shuttling rates only affect $\Erk1p_{\n}$ shuttling and Erk2p shuttling only affect $\Erk2p_{\n}$ steady state. By increasing and decreasing the Erk2p shuttling rates, we find a quantitatively larger difference in steady state value of $\Erk2p_{\n}$. This difference confirms why Erk2p shuttling is most important in the complete system. By examining the effects of Mek shuttling on $\Erk1/2p_{\n}$ steady states, it is clear that $\Erk2p_{\n}$ responds more than $\Erk1p_{\n}$ both in activation time and steady state value. While it is well known that $\Mek$ has a nuclear exclusion sequence that exports it from the nucleus, studies have suggested that $\Mek$ may play an active role in transport of $\Erk$ \cite{Adachi:2000vt,Adachi:1999ha}. Our findings that the shuttling of $\Mek$ is less sensitive than Erk1p/Erk2p, yet result in the same effect on steady state and activation time (compare Fig.~6b and Fig.~6c) is somewhat counterintuitive. Evidence suggests that the Mek mediated contribution to Erk export is at most marginal under physiological conditions, only contributing to the steady-state in unphosphorylated conditions \cite{costa:jcs:2006}. This may be due to the very low rate supported by the CRM1 mediated export. Despite the numerous studies to understand the role Mek plays in regulating Erk, our analysis gives a juxtaposition: Mek shuttling sensitivity is relatively small compared to Erk1p/Erk2p shuttling sensitivity while simulations predict that $\Mek$ and $\Erk$ shuttling have the same effect on activated nuclear $\Erk$.

The effects of dephosphorylation on phospho-nuclear $\Erk$ steady-states is more pronounced than the effect of phosphorylation rates. By varying the dephosphorylation rates of Erk2 ($k_{12} = 1.4~\times~10^{-1}$ baseline) a decrease by an order of magnitude remarkably affects the steady state value of $\Erk2p_{\n}$. Increasing the dephosphorylation rate more than four-fold ($k_{12} = 6~\times~10^{-1}$ ) decreases the steady state approximately half of the baseline; however the potential for increasing the activation time and steady state value by changing the dephosphorylation confirm the findings from the minimal model. By studying the effects of phosphorylation rates on the steady-state values, the phosphorylation rates do not exhibit a notable change in either activation or steady-state, which is a significant difference between the minimal and complete models.

\begin{figure}
\begin{center}
\includegraphics[width=\textwidth]{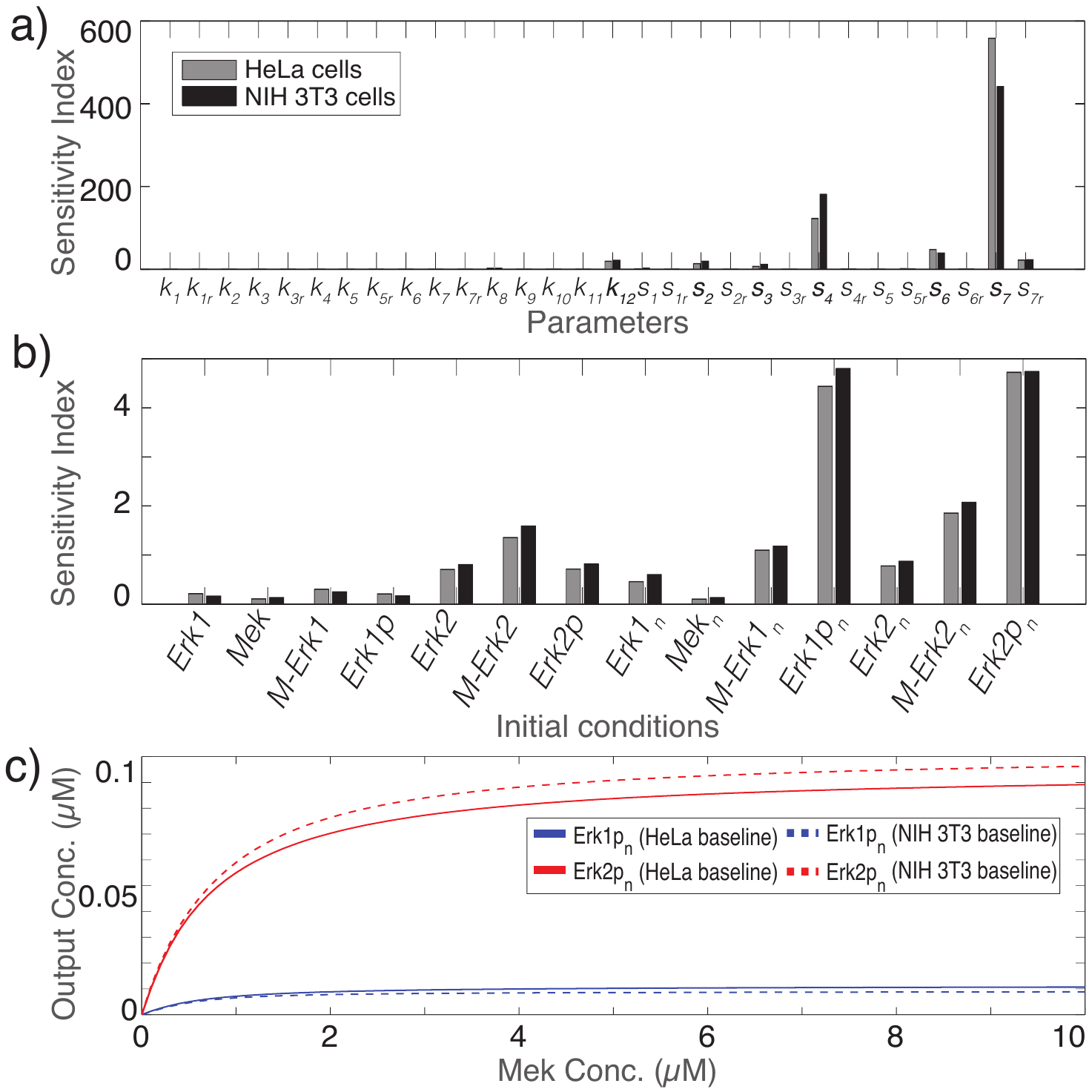}
\caption{Analysis of complete system. Sensitivity coefficients for parameters (a) and initial conditions (b). For both HeLa and NIH cells only a limited number of parameters have a significant influence on the behavior of $\Erk1p_{\n}$ and $\Erk2p_{\n}$: $s_{7}, s_{4}, s_{6}, s_{7r}, k_{12}, s_{2},s_{3}$.
Sensitivity to initial conditions reveals that for both cell types all initial conditions have a comparable impact on the observed dynamics. The most sensitive initial conditions, however, are much less sensitive than most sensitive parameters. (c) Effects of phosphorylated Mek on output concentrations $\Erk1p_{\n}$ and $\Erk2p_{\n}$.}
\end{center}
\end{figure}

The complete system is consistent with the main finding that the shuttling rates of phosphorylated $\Erk$ affect the steady-state values of $\Erk1p_n$ and $\Erk2p_n$ (Fig.~5a). The sensitivity analysis for HeLa cells suggests it is more sensitive to Erk2p shuttling whereas NIH 3T3 cells are more sensitive to Erk1p shuttling; however our simulations do not show remarkable differences between the two cell types. We also observe that both cell types are responsive to Mek shuttling speeds although Mek shuttling parameters are not overly sensitive in analysis, which may explain the ongoing investigations of the role of Mek shuttling on Erk. 
\begin{figure}
\begin{center}
\includegraphics[width=4in]{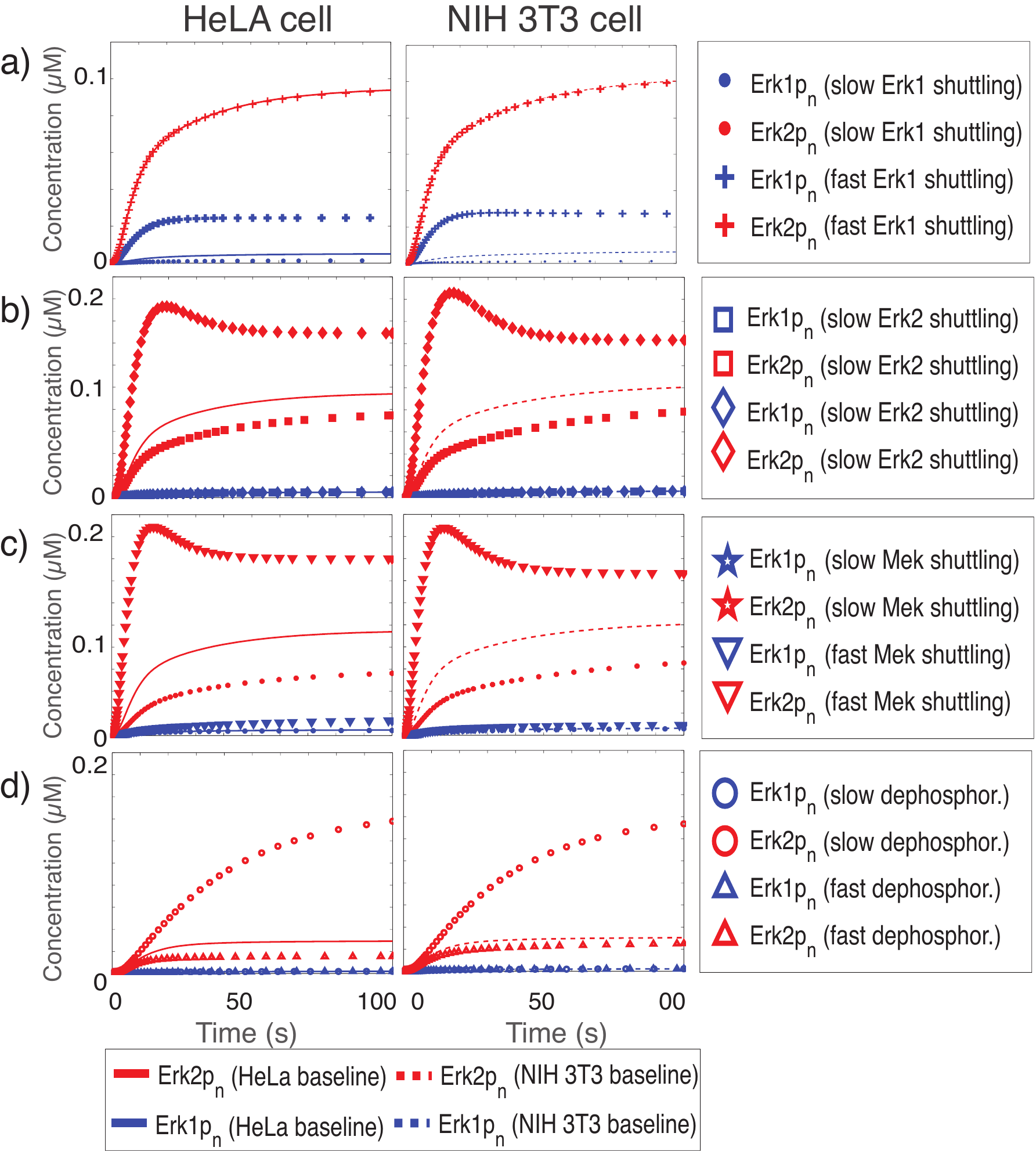}
\caption{Effects of Erk and Mek shuttling, and dephosphorylation rates for complete model at baseline parameter values and initial conditions. Solid line denotes HeLa cell baseline and dotted line denotes NIH 3T3 cell baseline. a,b) Effect of Erk1p and Erk2p shuttling speeds. The shuttling parameters, $s_4$ and $s_7$, are varied by an order of magnitude from their baseline values. The solution curves for varying Erk1p shuttling rates is given by slow (dots) and fast (cross) whereas Erk2p shuttling rates is given by slow (squares) and fast (diamonds) can be compared to their respective baseline values. c) Effects of Mek shuttling rates. The Mek import shuttling parameters, $s_2$, $s_3$ and $s_6$, are varied from their baseline values. The time course for slow $\Mek$ shuttling is denoted by stars, fast $\Mek$ shuttling is represented by upside-down diamonds lines. d) Effects of dephosphorylation rate of Erk2p. The phosphorylation parameter, $k_{12}$ is varied from its baseline values. The solution curves for slow rates ($k_{12}=0.014$) are denoted by circles, fast rate ($k_{12}=0.6$) are represented by triangles. }
\end{center}
\end{figure}

\subsection*{Discussion}
Our analysis suggests that transport of $\Erk$ is most important for determining activity levels.  Competition between $\Erk1$ and $\Erk2$ for Mek is induced as biochemical rates and total concentrations of $\Erk$ and $\Mek$ change; our minimal model and analysis may also offer an explanation of why the Mek--Erk stochiometry is close to one.  Finally, analysis of the complete model highlights the unexpected result that as Mek shuttling rates are less sensitive in analysis, the simulation of the solution curves provides the same effects as varying the Erk2p shuttling rate.

The differences between cell types are also illuminating: since the shuttling rates are the most sensitive parameter one may be led to think that differences in trafficking are at the basis for heterogeneity. There are two additional points that we feel are pertinent in this context: (1) the composition and amount of nucleoporins changes during the cell cycle, which bring about changes in trafficking rates and in levels of Erk activation; (2) in neuronal cells, where trafficking rates are maximally different between Erk1 and Erk2 (unpublished data), we may expect enhanced difference between Erk1 and Erk2 activation levels. This is notable (it is certainly enthralling from a neurobiological point of view) since the only clear phenotype of the Erk1 knockout is neurophysiological. Although beyond the scope of this study, considering changes in transient changes in Erk isoform concentration as well as exploring the effects of downstream substrates may pin down the physiological implication of the shuttling mechanisms discussed here. 

In response to a given stimulus the different activation times are mostly due, not to differences in the phospho-dephospho rates, but to differences in shuttling speed. The FRAP experiments show that there is a large variability in trafficking speed, likely due to the variable density of nucleoporins, and this could go a long way towards justifying the different kinetics. In view of the large individual differences between cells, this seems to be interesting and hitherto neglected aspects of studies of cell-to-cell variability. Through the development of these abstractions and sensitivity analysis as performed might help to understand what are the most likely parameters responsible for the heterogeneous responses within a cell population. 

More generally, our results highlight the important aspect of spatial localization of specific isoforms that has hitherto often been neglected in models and experiments. Similarly to MAPK studies, NFAT has been found to have different behavior based on stimuli duration (sustained, pulsed etc); notably, NFAT isoforms in the Calcium/NFAT pathway that regulate T cells have different timed responses which may be explained by tissue specific subcellular localization (Nir Friedman, personal communications). These are not isolated cases: even Fox-1 homologs are tissue-dependent and a theoretical study of their nuclear localization, such as this one, may determine how these isoforms regulate activities of neural tissue-specific splicing. Thus we feel that differential shuttling of signaling molecule isoforms coupled to competition between them may have a general role on cellular decision making processes.

\section*{Acknowledgements}
HAH and MPHS gratefully acknowledge funding from Leverhulme Trust. MBD is
supported by a BBSRC-Microsoft Research Dorothy Hodgkin Postgraduate
Award. M.K. acknowledges support from the Biotechnology and Biological
Sciences Research Council (BBSRC) (BB/G020434/1). PRIN 2008 from MIUR to GMR.

\section*{Supplementary Material}
There are two sections which include all model equations, minimal model parameters and initial conditions and Fig.~S1.

\section*{Models}
\subsection*{Equations}
The biochemical reactions of the full system given in Table~1  are translated into a system of ODEs following mass action kinetics:
{\footnotesize
\begin{align*}
            d\conc{\Erk1} / dt &= -k_{1}\conc{\Erk1}\conc{\Mek}+k_{-1}\conc{\Merk1}+k_{9}\conc{\Erk1p} + s_{-1}\conc{\Erk1_n}- s_{1}\conc{\Erk1},   \\
            d\conc{\Mek} / dt &=  -k_{1}\conc{\Erk1}\conc{\Mek} - k_{5}\conc{\Erk2}\conc{\Mek} + (k_{-1}+k_{2})\conc{\Merk1} \\&+(k_{-5}+k_{6})\conc{\Merk2} + s_{-2}\conc{\Mek_n}- s_{2}\conc{\Mek1},  \\ 
            d\conc{\Merk1} / dt &=  k_{1}\conc{\Erk1}\conc{\Mek} - (k_{-1}+k_{2}) \conc{\Merk1} + s_{-3}\conc{\Merk1_n} - s_{3}\conc{\Merk1},  \\
            d\conc{\Erk1p} / dt &=  k_{2} \conc{\Merk1}-k_{9}\conc{\Erk1p} + s_{-4}\conc{\Erk1p_n} - s_{4}\conc{\Erk1p},   \\
            d\conc{\Erk2} / dt &= -k_{5}\conc{\Erk2}\conc{\Mek}  + k_{-5}\conc{\Merk2} + k_{11}\conc{\Erk2p} + s_{-5}\conc{\Erk2_n} - s_{5}\conc{\Erk2},  \\
            d\conc{\Merk2} / dt &= k_{5}\conc{\Erk2}\conc{\Mek}-(k_{-5}+k_{6})\conc{\Merk2}  + s_{-6}\conc{\Merk2_n} - s_{6}\conc{\Merk2},  \\
            d\conc{\Erk2p} / dt &= k_{6}\conc{\Merk2}-k_{11}\conc{\Erk2p} + s_{-7}\conc{\Erk2p_n} - s_{7}\conc{\Erk2p},  \\
            d\conc{\Erk1_n} / dt &= -k_{3}\conc{\Erk1_n}\conc{\Mek_n}+k_{-3}\conc{\Merk1_n}+k_{10}\conc{\Erk1p_n} + s_{1}\conc{\Erk1}- s_{-1}\conc{\Erk1_n},  \\
            d\conc{\Mek_n} / dt &=  -k_{3}\conc{\Erk1_n}\conc{\Mek_n} - k_{5}\conc{\Erk2_n}\conc{\Mek_n} + (k_{-3}+k_{4})\conc{\Merk1_n}\\& +(k_{-7}+k_{8})\conc{\Merk2_n} + s_{2}\conc{\Mek} - s_{-2}\conc{\Mek_n},\\
            d\conc{\Merk1_n} / dt &=  k_{3}\conc{\Erk1_n}\conc{\Mek_n} - (k_{-3}+k_{4}) \conc{\Merk1_n} +s_{3}\conc{\Merk1} - s_{-3}\conc{\Merk1_n}, \\
            d\conc{\Erk1p_n} / dt &=  k_{4} \conc{\Merk1_n}-k_{10}\conc{\Erk1p_n} +s_{4}\conc{\Erk1p} - s_{-4}\conc{\Erk1p_n},  \\
            d\conc{\Erk2_n} / dt &= -k_{7}\conc{\Erk2_n}\conc{\Mek_n}  + k_{-7}\conc{\Merk2_n} + k_{12}\conc{\Erk2p_n} +s_{5}\conc{\Erk2} - s_{-5}\conc{\Erk2_n}, \\
            d\conc{\Merk2_n} / dt &= k_{7}\conc{\Erk2_n}\conc{\Mek_n}-(k_{-7}+k_{8})\conc{\Merk2_n}  +s_{6}\conc{\Merk2} - s_{-6}\conc{\Merk2_n},  \\
            d\conc{\Erk2p_n} / dt &= k_{8}\conc{\Merk2_n}-k_{12}\conc{\Erk2p_n} +s_{7}\conc{\Erk2p} - s_{-7}\conc{\Erk2p_n}. \\
\end{align*} } \normalsize

Note, continuous shuttling is dependent on $\Mek$ activation \cite{costa:jcs:2006}; therefore, we assume that Mek activation is sustained in the system such that total active $\Mek$ is constant, 
$$\Mek_T=\Mek+\Mek_n+\Merk1+\Merk1_n+\Merk2+\Merk2_n.$$
Since this study only focusses on the short timescales of Erk1 and Erk2 activation and not ensuing gene expression processes, we assume that total Erk1 and Erk2 are also constant and satisfy the conservation relations,
\begin{align*}
\Erk1_T&=\conc{\Erk1}+\conc{\Erk1_n}+\conc{\Erk1p}+\conc{\Erk1p_n}+\conc{\Merk1}+\conc{\Merk1_n}, \\
\Erk2_T&=\conc{\Erk2}+\conc{\Erk2_n}+\conc{\Erk2p}+\conc{\Erk2p_n}+\conc{\Merk2}+\conc{\Merk2_n},
\end{align*} 
where $\Erk1_T$ and $\Erk2_T$ are total Erk1 and Erk2, respectively.   

\par
The reactions of the minimal model are given in Table~1 and the relevant reactions are identified by only cytoplasmic reactions, corresponding to the first seven differential equations in the full model system, except the terms with a coefficient of $s_j$ are disregarded. Thus the model in Eqs$.~1-7$ become: {\footnotesize
\begin{align}
d\conc{\Erk1} / dt &= -k_{1}\conc{\Erk1}\conc{\Mek}+k_{-1}\conc{\Merk1}+k_{9}\conc{\Erk1p},  \\
d\conc{\Mek} / dt &=  -k_{1}\conc{\Erk1}\conc{\Mek} - k_{5}\conc{\Erk2}\conc{\Mek} \\&+ (k_{-1}+k_{2})\conc{\Merk1} +(k_{-5}+k_{6})\conc{\Merk2},  \\ 
d\conc{\Merk1} / dt &=  k_{1}\conc{\Erk1}\conc{\Mek} - (k_{-1}+k_{2}) \conc{\Merk1} , \\
d\conc{\Erk1p} / dt &=  k_{2} \conc{\Merk1}-k_{9}\conc{\Erk1p},   \\
d\conc{\Erk2} / dt &= -k_{5}\conc{\Erk2}\conc{\Mek}  + k_{5}\conc{\Merk2} + k_{11}\conc{\Erk2p},  \\
d\conc{\Merk2} / dt &= k_{5}\conc{\Erk2}\conc{\Mek}-(k_{5}+k_{6})\conc{\Merk2},  \\
d\conc{\Erk2p} / dt &= k_{6}\conc{\Merk2}-k_{11}\conc{\Erk2p},  
\end{align}}
and conservation relations:
\begin{align*} 
\Mek_T&=\conc{\Merk1}+\conc{\Merk2}+\conc{\Mek}, \\ 
\Erk1_T&=\conc{\Erk1}+\conc{\Erk1p}+\conc{\Merk1}, \\
\Erk2_T&=\conc{\Erk2}+\conc{\Erk2p}+\conc{\Merk2},
\end{align*} 
where $\Mek_T$, $\Erk1_T$ and $\Erk2_T \in \field{R}$ are total Mek, Erk1 and Erk2, respectively. We introduce the normalised variables $[\widetilde{\Erk1}] = [\Erk1]/\Erk1_T $, $[\widetilde{\Erk1p}] = [\Erk1p]/\Erk1_T$, $[\widetilde{\Erk2}] = [\Erk2]/\Erk2_T$, and $[\widetilde{\Erk2p}] = [\Erk2p]/\Erk2_T$. Using the conservation relations above and the new normalised variables
we obtain the following equations: 
\begin{align*}
d\conc{\widetilde{\Erk1}} / dt &= -k_{1}\Erk1_T\conc{\widetilde{\Erk1}}\left( \frac{\Mek_T}{\Erk1_T} - \mathcal{M}_1 - \mathcal{R}  \mathcal{M}_2 \right)+k_{-1}\mathcal{M}_1+k_{9}\conc{\widetilde{\Erk1p}},  \\
d\conc{\widetilde{\Erk1p}} / dt &=  k_{2} \mathcal{M}_1-k_{9}\conc{\widetilde{\Erk1p}},   \\
d\conc{\widetilde{\Erk2}} / dt &= -k_{5}\Erk2_T\conc{\widetilde{\Erk2}}\left( \frac{\Mek_T}{\Erk2_T} - \mathcal{R}^{-1} \mathcal{M}_1 -  \mathcal{M}_2 \right)  + k_{5}\mathcal{M}_2 + k_{11}\conc{\widetilde{\Erk2p}} , \\
d\conc{\widetilde{\Erk2p}} / dt &= k_{6}\mathcal{M}_2-k_{11}\conc{\widetilde{\Erk2p}},  
\end{align*}
where
$$\mathcal{M}_1=1-\conc{\widetilde{\Erk1}}-\conc{\widetilde{\Erk1p}}, \qquad \mathcal{M}_2=1-\conc{\widetilde{\Erk2}}-\conc{\widetilde{\Erk2p}}, \qquad \textrm{and} \qquad \mathcal{R} = \frac{\Erk2_T}{\Erk1_T},$$
and the system has been reduced by three equations.
For simplicity, in the main section of the article, we drop the tilde notation from the variables. 

\subsection*{Estimation of transport parameters and initial conditions}
Before photobleaching, the cells are assumed to be in equilibrium and that influx and efflux of the Erk species are equal.  After bleaching, the flux continues in and out of the cell, and fluorescence is measured. We introduce two equations to describe the bleached nuclear Erk species ($F_{\n}^B$) and recovered fluorescence of nuclear Erk species ($F_{\n}$):   $$\frac{d[F_{\n}^B]}{dt} = -s_{-j} F_{\n}^B \quad\textrm{and}\quad \frac{d[F_{\n}]}{dt} = \alpha-s_{-j} F_{\n}$$ where $s_{-j}$ is the rate at which bleached species leaves the nucleus and $\alpha$ is the rate at which unbleached, fluorescent species enters the nucleus. At equilibrium ($\frac{d[F_n]}{dt} = 0 = \alpha- s_{-j} [F_{\n}]$), we have $[F_n] = \alpha / s_{-j}.$ As before, we normalize by the steady-state of $F_n$ ($\alpha / s_{-j}$): $\widetilde{F_n} = \frac{[F_n]}{\alpha/ s_{-j}}$, giving $[F_n] = \frac{\alpha}{s_{-j}}\widetilde{F_n}$, so at equilibrium $\widetilde{F_n}=1$, and $\frac{d\widetilde{F_n}}{dt} = s_{-j} - s_{-j} \widetilde{F_n}.$ Then the solution is $\widetilde{F_n} = 1-e^{-t \cdot s_{-j}}$. If at $t = \tau$, and $F^{*}$ is the measured fluorescence $(\widetilde{F_n} = F^{*})$, then $F^{*} = 1-e^{-\tau \cdot s_{-j}}$ and the nuclear export rate is given by $s_{-j} = -\frac{\ln(1-F^{*})}{\tau}$. Here, the $F^{*}$ is the recovered fluorescence at the time constant $\tau$. In NIH 3T3 cells we have for starved Erk1 $\tau=653$s, stimulated $\Erk1p$ $\tau =266$s, starved Erk2 $\tau=178$s,  and stimulated $\Erk2p$ $\tau =84$s when $F^{*} \approx 0.75$ normalized fluorescence \cite{marchi:po:2008}.  Given that the sizes of HeLa and NIH 3T3 cells are similar (the ratio of molecules in cytoplasm/nucleus is $1.5\pm0.2$ for $\Erk$ and $1.1\pm0.72$ for $\Erkp$) the import rates ($s_j$) can be estimated as export rate divided by the molecular number ratio, as calculated by \cite{fujioka:jbc:2006}. Without the shuttling rates of Mek for NIH 3T3 cells, we assume the values as calculated by \cite{fujioka:jbc:2006} for HeLa cells.   

Initial conditions of the transport model are the same as the full model. The initial conditions of the biochemical model follow the normalization such that Erk1(0)=1, Erk2(0)=1 and Mek(0)=1, and the total concentrations of Erk1 ($\Erk1_T$) and Erk2 ($\Erk2_T$) are set to maintain that Erk2 is four times more abundant than Erk1.    

\section*{Total Nuclear Erk}
  \paragraph*{Supplementary Fig.~1}
Time course of total nuclear Erk for HeLa and NIH 3T3 cells A. Full model. (\rmnum{1}) Time course of total nuclear $\Erk1$ ($\Erk1p_n +\Erk1_n$) and $\Erk2$ ($\Erk2p_n +\Erk2_n$) at baseline conditions.  (\rmnum{2}) Time course of total nuclear $\Erk1$ ($\Erk1p_n +\Erk1_n$) and $\Erk2p$ ($\Erk2p_n +\Erk2_n$) at equal Erk1/2 ($\Erk1=\Erk2=0.5$). (\rmnum{3}) Time course of total nuclear $\Erk1$ ($\Erk1p_n +\Erk1_n$) and $\Erk2$ ($\Erk2p_n +\Erk2_n$) at Limited Mek ($\Mek=0.2$). B) Transport model. Transport model was simulated at baseline parameter values for HeLa cells (solid) and NIH 3T3 cells (dashed).  

 {\includegraphics[width=\textwidth]{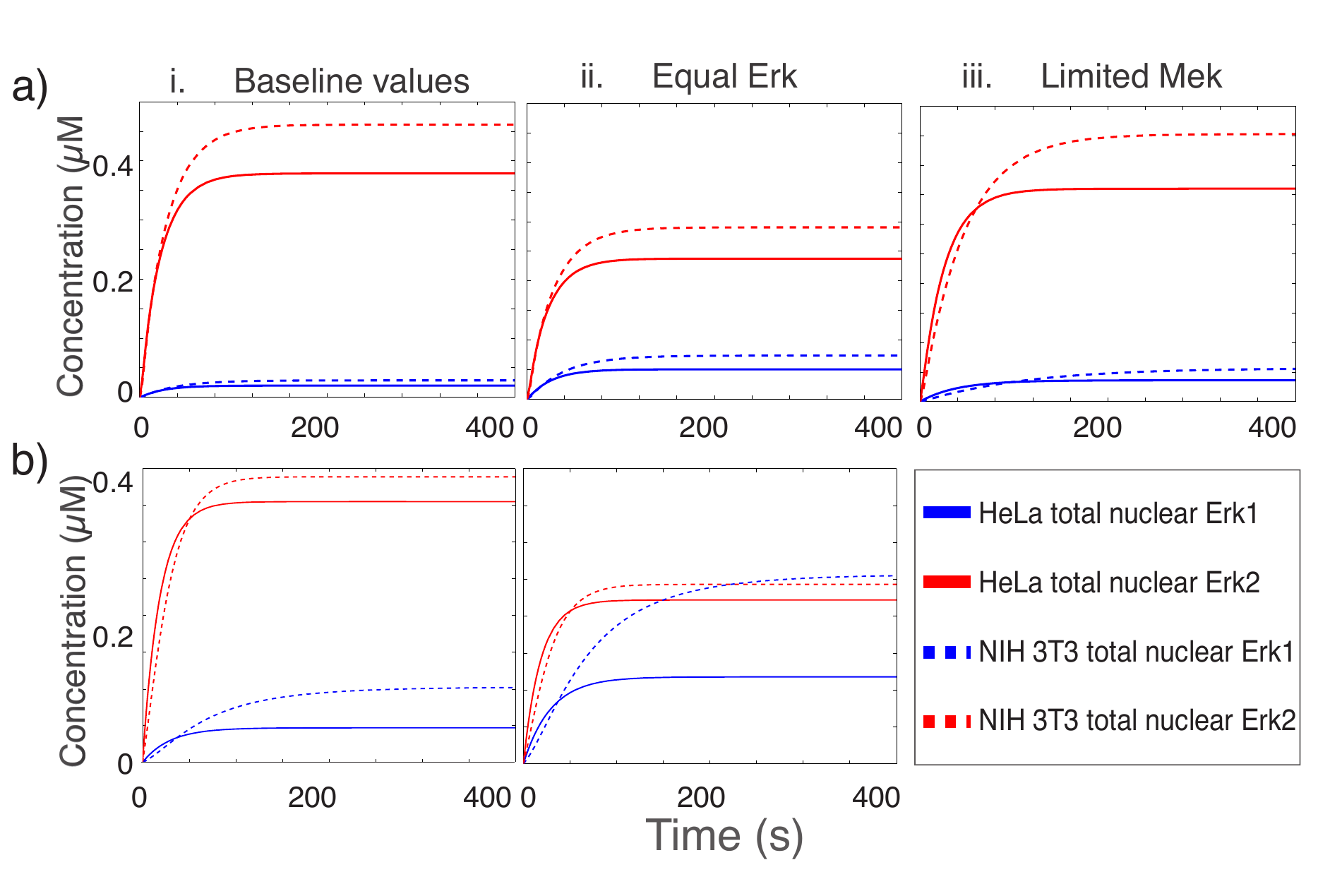}

\bibliographystyle{Vancouver}
\bibliography{ErkBib21June2011}

\end{document}